\pdfoutput=1
\documentclass[twocolumn]{aastex63}
\usepackage{natbib}
\usepackage{hyperref}
\usepackage{lineno}
\usepackage{ulem}

\def\deg{\ifmmode^\circ\else$^\circ$\fi}
\def\kps{km\thinspace s$^{-1}$}
\newcommand{\Msolar}{$M_{\odot}$}

\newcommand{\simi}{$\sim$}

\newcommand{\htwo}{H\,{\sc ii}}

\newcommand{\grstco}{$^{13}$CO($J$ = 1--0)}
\newcommand{\chmstco}{$^{13}$CO($J$ = 3--2)}
\pdfoutput=1

\shorttitle{Physical processes in the IRDC hosting N59-North}
\shortauthors{A.~K. Maity et al.}

\begin{document}
\title{Investigating Embedded Structures and Gas Kinematics in the IRDC Hosting Bubble N59-North}

\correspondingauthor{A.~K. Maity}
\email{Email: aruokumarmaity123@gmail.com}

\author[0000-0002-7367-9355]{A.~K. Maity}
\affiliation{Astronomy \& Astrophysics Division, Physical Research Laboratory, Navrangpura, Ahmedabad 380009, India.}
\affiliation{Indian Institute of Technology Gandhinagar Palaj, Gandhinagar 382355, India.}

\author[0000-0001-6725-0483]{L.~K. Dewangan}
\affiliation{Astronomy \& Astrophysics Division, Physical Research Laboratory, Navrangpura, Ahmedabad 380009, India.}

\author[0009-0001-2896-1896]{O.~R. Jadhav}
\affiliation{Astronomy \& Astrophysics Division, Physical Research Laboratory, Navrangpura, Ahmedabad 380009, India.}
\affiliation{Indian Institute of Technology Gandhinagar Palaj, Gandhinagar 382355, India.}

\author[0000-0001-5731-3057]{Saurabh Sharma}
\affiliation{Aryabhatta Research Institute of Observational Sciences, Manora Peak, Nainital 263002, India.}

\author[0000-0002-6740-7425]{Ram Kesh Yadav}
\affiliation{National Astronomical Research Institute of Thailand (Public Organization), 260 Moo 4, T. Donkaew, A. Maerim, Chiangmai 50180, Thailand}

\author[0000-0002-8966-9856]{Y.~Fukui}
\affiliation{Department of Physics, Nagoya University, Furo-cho, Chikusa-ku, Nagoya 464-8601, Japan}

\author[0000-0003-2062-5692]{H.~Sano}
\affiliation{Faculty of Engineering, Gifu University, 1-1 Yanagido, Gifu 501-1193, Japan}

\author[0000-0002-7935-8771]{T.~Inoue}
\affiliation{Department of Physics, Nagoya University, Furo-cho, Chikusa-ku, Nagoya 464-8601, Japan}
\affiliation{Department of Physics, Konan University, Okamoto 8-9-1, Kobe, Japan}

\begin{abstract}
We present a multi-wavelength study of an extended area hosting the bubble N59-North to explore the physical processes driving massive star formation (MSF). The {\it Spitzer} 8 $\mu$m image reveals an elongated/filamentary infrared-dark cloud (length $\sim$28 pc) associated with N59-North, which contains several protostars and seven ATLASGAL dust clumps at the same distance. The existence of this filament is confirmed through $^{13}$CO and NH$_3$ molecular line data in a velocity range of [95, 106] km s$^{-1}$. All dust clumps satisfy Kauffmann \& Pillai's condition for MSF. Using {\it Spitzer} 8 $\mu$m image, a new embedded hub-filament system candidate (C-HFS) is investigated toward the ATLASGAL clump, located near the filament's central region. MeerKAT 1.3 GHz continuum emission, detected for the first time toward C-HFS, reveals an ultracompact {\htwo} region driven by a B2-type star, suggesting an early stage of HFS with minimal feedback from the young massive star. The comparison of the position-velocity (PV) and position-position-velocity (PPV) diagrams with existing theoretical models suggests that rotation, central collapse, and end-dominated collapse are not responsible for the observed gas motion in the filament. The PPV diagram indicates the expansion of N59-North by revealing blue- and red-shifted gas velocities at the edge of the bubble. Based on comparisons with magnetohydrodynamic simulations, this study suggests that cloud-cloud collision (CCC) led to the formation of the filament, likely giving it a conical structure with gas converging toward its central region, where C-HFS is located. Overall, the study supports multi-scale filamentary mass accretion for MSF, likely triggered by CCC.
\end{abstract}
%
%
\section{Introduction}
\label{sec:intro}
Massive OB stars ($>$ 8 $M_{\odot}$) continuously emit radiative and mechanical energy, significantly shaping the surrounding environment throughout their existence. However, despite their dominant impact, the processes underlying their origin and feedback remain poorly understood and are active areas of research \citep{zinnecker07,tan14,Motte+2018,rosen20}. Earlier studies \citep[e.g.,][]{myers09,kumar20,liu23,yang23} have established that Hub-filament systems (HFSs), which are key structures in the molecular clouds, play a central role in star formation, particularly in the birth of massive stars and star clusters. It is worth noting that the impact of massive OB stars on their parent molecular environment can modify the initial conditions for massive star formation (MSF) and reshape the early configuration of HFSs. Therefore, studying the early stages of HFSs, where the impact of massive stars is minimal, offers valuable insights into the initial conditions for the development of hubs and their massive star-forming activity \citep[e.g.,][]{Dewangan_2024MNRAS}. The Galactic `Snake,' an infrared-dark cloud (IRDC) G11.11-0.12 illustrates this phase, characterized by unperturbed filamentary structures and a compact central hub \citep{Dewangan_2024MNRAS,Bhadari_2025_A&AL}. 
In contrast, an evolved HFS represents a later stage in the HFS lifecycle, defined by active feedback from massive stars. Mon R2 (RA(J2000) = 06$^{\rm h}$07$^{\rm m}$46.2$^{\rm s}$, Dec(J2000) = $-$06$\deg$23$'$8$\rlap.{''}3$) serves as an example of such an evolved HFS \citep[][]{morales19,dewangan24xb}. Although the early stages of HFSs are important for understanding the initial conditions of MSF, studies on such sites remain scarce in the literature. 

This study focuses on a large area encompassing the mid-infrared (MIR) bubble N59 \citep{churchwell06,deharveng10,hattori16,hanaoka19} and its northern edge, known as N59-North \citep[][]{chen24}. 
The bubble N59 is located toward the Galactic coordinates ({\it l}, {\it b}) = (33\rlap.$\deg$071, $-$0\rlap.$\deg$075) and exhibits a broken ring or shell-like structure in both molecular line data and dust continnum emission \citep[e.g.,][]{paulson24}. 
\citet{Anderson_2009ApJ} estimated a kinematic distance of 5.6$\pm$2 kpc for N59. However, using Gaia parallax measurements \citep{gaia16,gaia21}, \citet{paulson24} refined this to a more accurate distance of 4.66$\pm$0.70 kpc. It is important to note that Gaia parallax measurements beyond 2 kpc lead to systematic underestimation of distances depending on the fractional parallax uncertainty. However, when this uncertainty remains below 0.2, distance estimates can still be considered reliable up to 10 kpc. More details about the distance estimates can be found in \href{https://gea.esac.esa.int/archive/documentation/GDR3/Data_analysis/chap_cu8par/sec_cu8par_validation/ssec_cu8par_qa_distances.html}{Gaia data release 3 documentation}. \citet{paulson24} selected sources with fractional parallax uncertainties below 0.1, ensuring a reliable distance estimate for this target site, which we adopt in this study.
The N59-North region contains several Class II 6.7 GHz methanol maser emissions (MMEs) and ultracompact (UC) {\htwo} regions, which serve as the indicators of early stages of MSF \citep[e.g.,][]{deharveng10,paulson24}. Using a multi-wavelength approach, \citet{paulson24} identified a HFS candidate associated with N59-North. However, additional studies are needed to validate the existence of this proposed HFS candidate.

Using Milky Way Imaging Scroll Painting (\href{http://english.dlh.pmo.cas.cn/ic/in/}{MWISP}) $^{12}$CO($J$ = 1--0) and $^{13}$CO($J$ = 1--0) line data, \citet{chen24} reported multiple velocity components toward the N59 bubble, specifically [65, 79], [79, 86], [86, 95], and [95, 108] km s$^{-1}$ (see Figure~2 in their work). Based on the detection of different velocity components, they suggested that multiple collision events have occurred among these velocity components over the past 2 Myr. However, the physical connections between these velocity components and their respective distances are not thoroughly addressed in their study. Among these, the velocity component in the range of [95, 108] km s$^{-1}$ appears to correspond to the major cloud associated with N59/N59-North, exhibiting a filamentary structure \citep[see Figure~6 in][]{chen24}. Despite the availability of extensive observational datasets and several studies focusing on N59/N59-North, the full extent of this filamentary structure remains unexplored, pointing to physical processes that are yet to be fully investigated.
In Figure~\ref{fg1}a, we highlight the extended IRDC, visible in the {\it Spitzer} 8.0 $\mu$m image (indicated by arrows), appearing as a filamentary structure in absorption. The locations of N59 and N59-North are also indicated in Figure~\ref{fg1}a \citep[see also][]{chen24}. In general, IRDCs are known to host dense, cold molecular gas and dust that block  infrared (IR) radiation from background sources. They are regarded as important sites for studying the earliest phases of star formation, particularly in the context of MSF \citep[e.g.][and references therein]{Ragan_2009ApJ}.
Interestingly, the IRDC hosting N59-North 
has not been the focus of any previous studies. As a result, the formation and evolution of the filament, along with its associated massive star-forming activity, remain unexplored. In this study, we employ a multi-wavelength observational approach to investigate the physical environment and star formation processes in the IRDC. We present a detailed kinematic analysis of the structures embedded in the IRDC using {\grstco}, {\chmstco}, and NH$_3$(1--1) line data. 

The outline of this paper is as follows: In Section~\ref{sec_data_N59}, we describe the observational data sets utilized in this work. The results derived from these data sets are detailed in Section~\ref{sec_result_N59}. The significance of these results related to MSF in the IRDC is discussed in Section~\ref{sec_disc_N59}. Finally, Section~\ref{sec_conc_N59} summarizes the key results and conclusions of the study.
\section{Data sets}
\label{sec_data_N59}
In this paper, we utilized multi-wavelength archival datasets (i.e., from near-infrared (NIR) to radio) as listed in Table~\ref{tab1}. 
The MAGPIS and SMGPS radio continuum data have rms noise levels ($\sigma$) of about 0.4 mJy beam$^{-1}$ and 20 $\mu$Jy beam$^{-1}$, respectively \citep{Helfand06,Goedhart_2024MNRAS}.
The {\grstco} line data from the Boston University-Five College Radio Astronomy Observatory GRS are calibrated in 
the antenna temperature ($T_{\rm{A}}$) scale.
The $\sigma(T_{\rm{A}})$, velocity separation between the channels, and angular resolution of the GRS {\grstco} line data are about 0.13 K, 0.21 {\kps}, and 46$''$, respectively \citep[see][for more details]{jackson06}.
The CHIMPS {\chmstco} line data \citep[having critical density $\geq 10^4$ cm$^{-3}$ at temperature $\leq$ 20 K;][]{rigby16} were also examined toward the selected target area. 
The $\sigma(T_{\rm{A}})$, the velocity separation between the channels, and the angular resolution of the CHIMPS {\chmstco} line data are about 0.6 K, 0.5 {\kps}, and 15$''$, respectively. To enhance the visibility of faint or diffuse features, the CHIMPS line data were smoothed using a Gaussian function, resulting in an angular resolution of {\simi}27$''$ and $\sigma(T_{\rm{A}})$ {\simi}0.15 K. In addition, we also examined the RAMPS NH$_3$(1--1) line data \citep[having critical density {\simi} 3 $\times$ 10$^3$ cm$^{-3}$;][]{Hogge_2018ApJS} in this study. The rms noise level, the velocity separation between the channels, and the angular resolution of the RAMPS NH$_3$(1--1) data are about 0.16 K, 0.2 {\kps}, and 32$''$, respectively. We smoothed the RAMPS line data using a Gaussian function with Full Width at Half Maximum (FWHM) of about 6$''$. 
In the direction of the selected target area, the ATLASGAL clumps at 870 $\mu$m \citep[from][]{urquhart18} and the positions of YSOs from the {\it Spitzer}/IRAC Candidate YSO (SPICY) Catalog \citep{kuhn21} were collected. 
The {\it Herschel} dust temperature ($T_\mathrm{d}$) and H$_2$ column density ($N(\rm H_2$)) maps (resolution $\sim$12$''$) were obtained from \citet{marsh17}. 
These maps were generated using the Bayesian PPMAP procedure \citep{marsh15}, applied to the {\it Herschel} 70--500 $\mu$m images from the Hi-GAL survey \citep{Molinari10b}. 
The photometric magnitudes of point-like sources at 3.6, 4.5, and 5.8 $\mu$m were obtained from the {\it Spitzer} GLIMPSE-I Spring $'$07 catalog \citep{benjamin03}. The data sets utilized in this work can be downloaded from the links provided in Section~\ref{sec_data_avail}.
\begin{table*}
  \tiny
\setlength{\tabcolsep}{0.05in}
\centering
\caption{A list of multi-wavelength surveys utilized in the present work.}
\label{tab1}
\begin{tabular}{lcccr}
\hline 
  Survey  &  Wavelength(s)/Frequency       &  Resolution ($\arcsec$)        &  Reference  \\   
\hline
\hline 
SARAO MeerKAT Galactic Plane Survey (SMGPS)                             & 1.3 GHz                       & $\sim$8          & \citet{Goedhart_2024MNRAS}\\
 Multi-Array Galactic Plane Imaging Survey (MAGPIS)                             & 20 cm                       & $\sim$6          & \citet{Helfand06}\\
 Galactic Ring Survey (GRS)                                                                   & 2.7 mm; $^{13}$CO($J$ = 1--0) & $\sim$46        &\citet{jackson06}\\
CO Heterodyne Inner Milky Way Plane Survey (CHIMPS)                                                                  & 0.9 mm; $^{13}$CO($J$ = 3--2) & $\sim$15        &\citet{rigby16}\\

 Radio Ammonia Mid-plane Survey (RAMPS) &  23.694 GHz; NH$_3$(1--1) & $\sim$32 & \citet{Hogge_2018ApJS}\\

APEX Telescope Large Area Survey of the Galaxy (ATLASGAL)                 &870 $\mu$m                     & $\sim$19.2        &\citet{schuller09}\\
{\it Herschel} Infrared Galactic Plane Survey (Hi-GAL)                              &70, 160, 250, 350, 500 $\mu$m                     & $\sim$5.8, $\sim$12, $\sim$18, $\sim$25, $\sim$37         &\citet{molinari10}\\
Inner Galactic plane survey using the Multiband Infrared Photometer for {\it Spitzer} (MIPSGAL)                                          &24 $\mu$m                     & $\sim$6         &\citet{carey05}\\ 
{\it Spitzer} Galactic Legacy Infrared Mid-Plane Survey Extraordinaire (GLIMPSE)       &3.6, 4.5, 5.8, 8.0  $\mu$m                   & $\sim$2, $\sim$2, $\sim$2, $\sim$2           &\citet{benjamin03}\\
\hline          
\end{tabular}
\end{table*}

%
\section{Results}
\label{sec_result_N59}
\subsection{Existence of an extended filamentary structure}\label{results_subsec1}
\begin{figure*}
\center
\includegraphics[width=10.7cm]{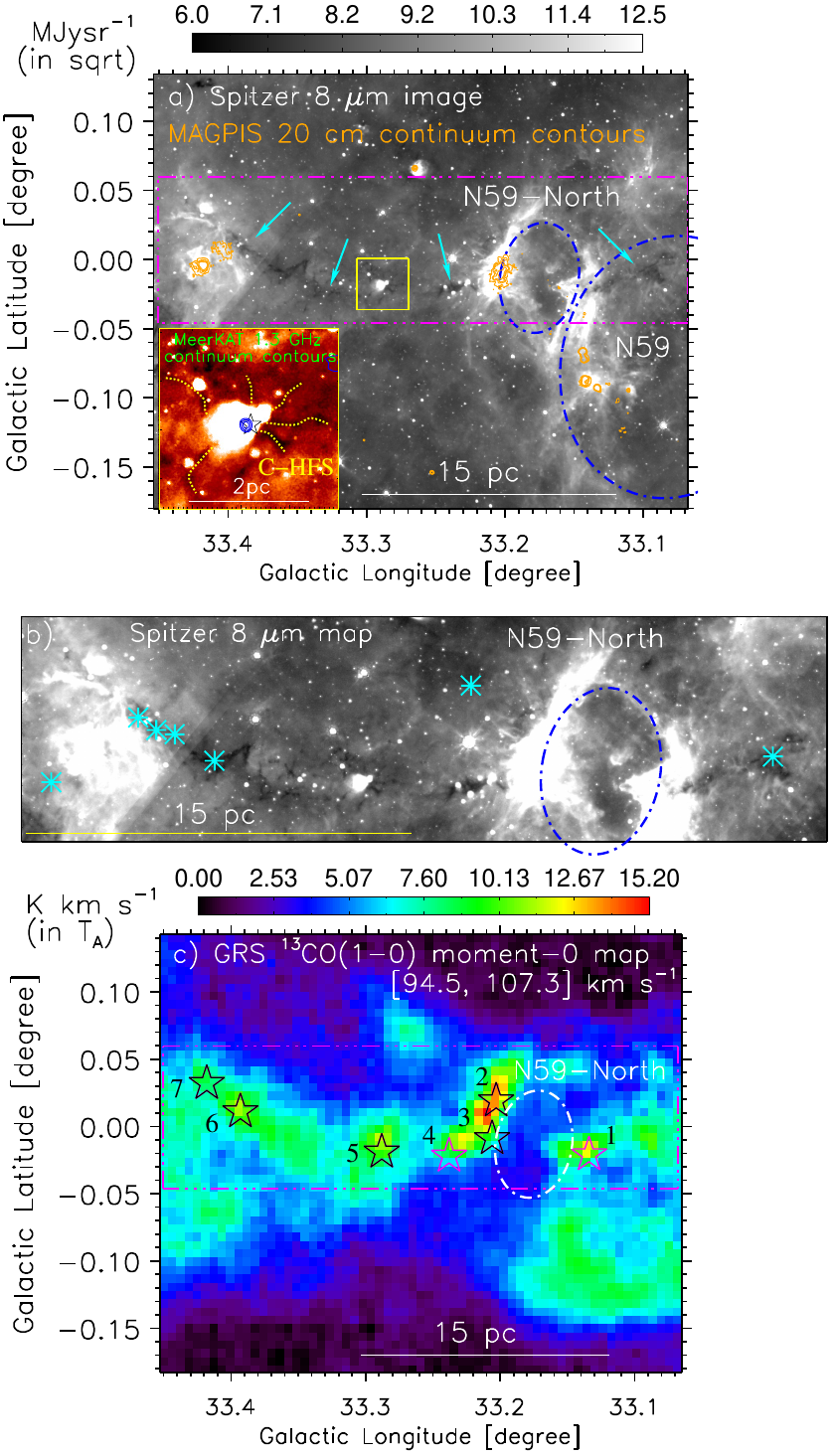}
\caption{(a) {\it Spitzer} 8 $\mu$m image of an area $\sim$0\rlap.$\deg$39 $\times$ 0\rlap.$\deg$31, centered at ({\it l}, {\it b}) = (33\rlap.$\deg$26, $-$0\rlap.$\deg$02), covering the IRDC associated with the bubble N59-North. The filamentary IRDC is marked with cyan arrows, while the bubbles N59 and N59-North are outlined by large and small ellipses, respectively. The orange contours present MAGPIS 20 cm radio continuum emission at levels [5, 8, 10 and 15] $\times \sigma$, with 1$\sigma$ $=$ 0.4 mJy beam$^{-1}$. The inset shows a zoomed-in view of the yellow rectangular region outlined over the IRDC. Blue contours indicate MeerKAT 1.3 GHz continuum emission at levels [5, 15 and 25] $\times \sigma$, with 1$\sigma = 20$ $\mu$Jy beam$^{-1}$. The position of the ATLASGAL clump (from \citet{urquhart18}) is shown with black star symbol. The IR-dark filaments are highlighted by yellow dotted lines in the inset. (b) Same image as panel ``a'' for the magenta dotted-dashed rectangular region. The position of the IRDC candidates toward the filament obtained from \citet{Pari_2020PASP} is marked using asterisk symbols. (c) GRS {\grstco} integrated intensity (moment-0) map for the velocity range of [94.5, 107.3] {\kps}. For the moment-0 map 1$\sigma$ = 0.21 K {\kps}. The black and magenta star symbols indicate the position of the ATLASGAL clumps with and without signature of outflow activity, respectively. 
The ellipse corresponding to N59-North is highlighted in panels ``b'' and ``c.'' A scale bar of 2 pc is provided in the inset of panel ``a,'' while a 15 pc scale bar is shown in each panel for a distance of 4.66 kpc.}
\label{fg1}
\end{figure*}

Figure~\ref{fg1}a presents the {\it Spitzer} 8.0 $\mu$m image overlaid with MAGPIS 20 cm continuum emission contours. 
As mentioned earlier, the {\it Spitzer} 8.0 $\mu$m image reveals an extended IRDC, as indicated by arrows. 
The MAGPIS radio continuum contours offer information on the distribution of ionized emission toward the IRDC. 
Previously reported structures, such as N59 and N59-North, are also labeled in the figure. 
%
Figure~\ref{fg1}b offers a closer view of this IRDC, including N59-North. Asterisks mark several IRDC candidates identified by \citet{Pari_2020PASP} through a semi-automated computational analysis of {\it Spitzer}/GLIMPSE data. This filamentary structure, as observed  in absorption, spans about 28 pc at the distance of 4.66 kpc and exhibits a disruption of nearly 3.8 pc at the location of the bubble N59-North. 
This extended filamentary structure has not been extensively studied to date. To confirm its existence as a single entity, we analyzed the GRS {\grstco} line data and generated an integrated intensity (moment-0) map over the velocity range of [94.5, 107.3] {\kps}. 
Figure~\ref{fg1}c displays the GRS moment-0 map, which reveals the molecular gas associated with the IRDC and highlights a similar filamentary morphology. 
The moment-0 map is also overlaid with the positions of the ATLASGAL 870 $\mu$m continuum 
clumps (see stars in Figure~\ref{fg1}c), situated at same distance ($\sim$6.5 kpc)
and exhibiting velocities 
between [99.2, 103.9] {\kps} \citep[from][]{urquhart18}. 
The alignment in velocity range and distance of the ATLASGAL clumps along the filamentary structure confirm its existence as a single physical entity, despite apparent disruptions near N59-North.

\begin{figure}
\center
\includegraphics[width=8cm]{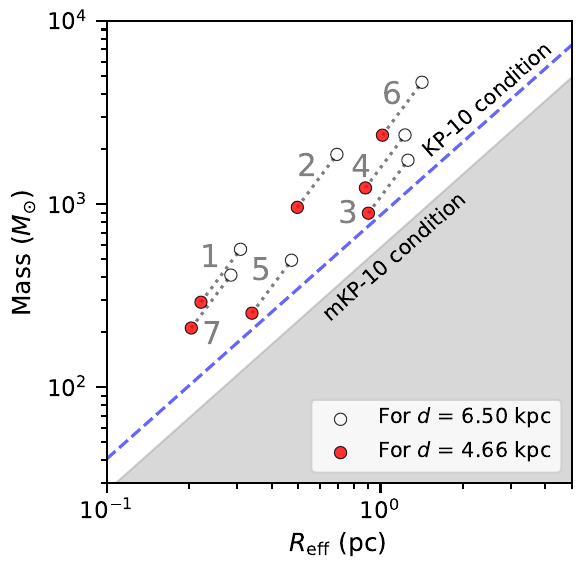}
\caption{The mass--effective radius plot for the ATLASGAL clumps. The data from \citet{urquhart18} for {\it d} = 6.5 kpc are displayed as open circles, while the red solid circles represent the data points scaled to our adopted distance of 4.66 kpc. The KP-10 condition for MSF is indicated by the blue dotted line. The mKP-10 condition for MSF as defined by \citet{urquhart18} corresponds to the white region above the gray-shaded area.}
\label{fg_ATL}
\end{figure}

\citet{Yang2018ApJS} studied the outflow activity associated with the ATLASGAL clumps using the CHIMPS $^{13}$CO/C$^{18}$O($J = 3$--$2$) line data. Clumps exhibiting outflow activity are indicated by black stars in 
Figure~\ref{fg1}c. No outflow activity was detected by \citet{Yang2018ApJS} for the ATLASGAL clump at $l = 33\rlap.{^\circ}134$. The C$^{18}$O($J = 3$--$2$) emission observed toward the clump at $l = 33\rlap.{^\circ}238$ was insufficient to search outflow activity. These two clumps are marked in magenta in Figure~\ref{fg1}c. The physical paremeters of these ATLASGAL clumps are listed in Table~\ref{tab_atlasgal_N59}. Notably, the ATLASGAL clump located at $l = 33\rlap.{^\circ}288$ (hereafter, ATL-5) is situated at the center of the filamentary cloud. 
To determine whether these clumps satisfy the empirical mass–size criterion for MSF, we have plotted the mass of the ATLASGAL clumps against their effective radius (see Figure~\ref{fg_ATL}). In Figure~\ref{fg_ATL}, the blue dashed line represents the Kauffmann \& Pillai condition for MSF, 
expressed as $M(R) = 870$ {\Msolar} $({R}/{\mathrm{pc}})^{1.33}$ \citep{Kauffmann_2010ApJ}. 
\citet{urquhart18} adjusted the mass coefficient by a factor of 1.5 to account for the difference in dust absorption coefficient used in their study. 
This modification yields the modified Kauffmann \& Pillai criterion (referred to as mKP-10) for MSF, defined as $M(R) > 580$ {\Msolar} $({R}/{\mathrm{pc}})^{1.33}$ \citep[see][]{urquhart18}. 
The white area above the gray-shaded region in the plot represents clumps that satisfy the mKP-10 condition. It is important to note that all ATLASGAL clumps meet the mKP-10 criterion for MSF, even after their masses and effective radii are adjusted to the distance of 4.66 kpc. 
\subsection{Investigation of a HFS candidate, C-HFS} 
\label{results_subsec2}
Figure~\ref{fg_ATL} clearly demonstrates that ATL-5 meets the empirical mass--size criterion for MSF. 
However, no MAGPIS 20 cm continuum emission is detected toward this clump. 
The bottom left inset in Figure~\ref{fg1}a is the {\it Spitzer} 8.0 $\mu$m image focusing on the central part of the filamentary structure in absorption (outlined by the solid box in Figure~\ref{fg1}a), where ATL-5 is located. 
Overlaid MeerKAT 1.3 GHz continuum emission contours within the inset reveal ionized gas concentrated around ATL-5, providing evidence of ongoing MSF activity within the clump. The detection of radio continuum emission is attributed to the higher sensitivity of the MeerKAT data compared to the MAGPIS data. We calculated the total flux density of the UC {\htwo} region (radius $\sim$0.05 pc) from MeerKAT data above the 5$\sigma$ threshold to be about 0.37 mJy. 
Using this value and assuming the temperature to be about 10$^4$ K, we determined the total number of Lyman continuum photons emitted per second (denoted as \(N_{\rm{UV}}\)), following the equation presented in \citet{Matsakis_1976AJ}.
We found that log(\(N_{\rm{UV}}\)) \(\sim\)44.8 for the driving source of the {\htwo} region. Comparing our calculated value of \(N_{\rm{UV}}\) with the theoretical estimation of \citet{Panagia_1973}, we infer that the source responsible for ionizing the gas is a B2-type star.


In the central part of the filamentary structure hosting N59-North, 
several pc-scale IR-dark filaments are identified in the {\it Spitzer} 8.0 $\mu$m image. These filaments seem to converge toward a common junction associated with ATL-5 (see the inset and the solid box in Figure~\ref{fg1}a). 
We estimated the background/foreground emission in the {\it Spitzer} 8 $\mu$m image using the median filtering technique \citep[see][for details]{Simon_2006ApJ,Ragan_2009ApJ} and found it to be about 70 MJy sr$^{-1}$ toward ATL-5. In contrast, the IR-dark filaments associated with ATL-5 are detected at levels of about 55--65 MJy sr$^{-1}$. These values indicate that the IR-dark filaments are not random background fluctuations but reliable features, positioning ATL-5 at the center of a HFS candidate (i.e., C-HFS) with an extent of less than 3 pc. 
The central hub (or ATL-5) is associated with an UC {\htwo} region.
The identification of the C-HFS is a new and important result, highlighting the potential of {\it Spitzer} 8 $\mu$m image in revealing pc-scale HFSs. Previously, \citet{Dewangan_2024MNRAS} identified multiple such HFSs using absorption features in the {\it Spitzer} 8 $\mu$m image toward the IRDC G11.11$-$0.12 \citep[see also][]{Bhadari_2025_A&AL}.
%
\subsection{Study of {\grstco}, {\chmstco}, and NH$_3$(1--1) data in the filamentary cloud}
\label{moment_mapol}
We investigate the gas distribution and kinematics in the filamentary cloud using the the GRS {\grstco}, CHIMPS {\chmstco} and RAMPS NH$_3$(1--1) line data.
\subsubsection{Molecular gas morphology and velocity in the filamentary cloud}
\label{moment_map_N59}
\begin{figure*}
\center
\includegraphics[width=0.7\textwidth]{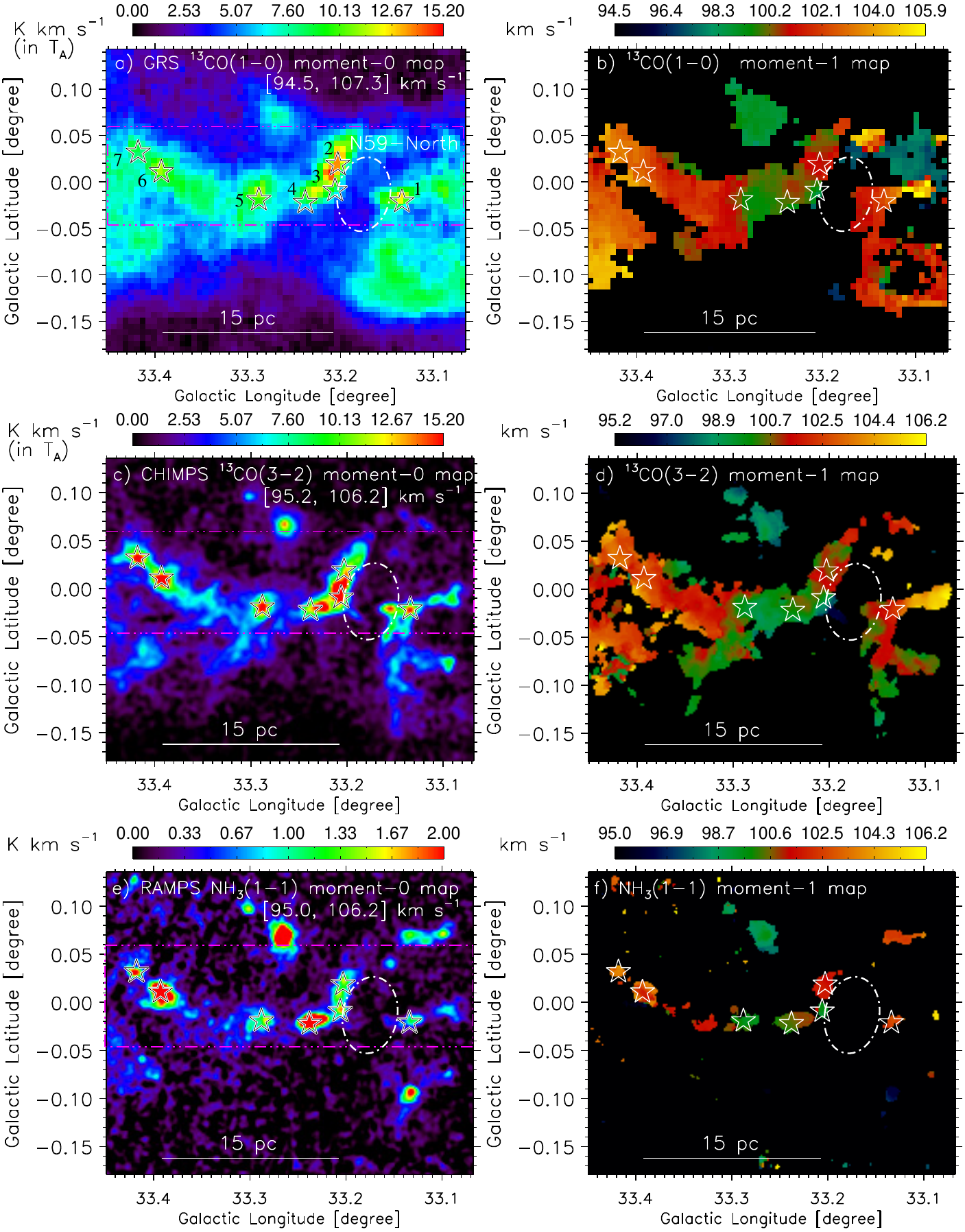}
\caption{(a) GRS {\grstco} moment-0 map, identical to Figure~\ref{fg1}c. (b) GRS {\grstco} intensity-weighted velocity (moment-1) map. (c) CHIMPS {\chmstco} moment-0 map for the velocity range of [95.2, 106.2] {\kps} (1$\sigma$ = 0.36 K {\kps}). (d) CHIMPS {\chmstco} moment-1 map. (e) RAMPS NH$_3$(1--1) moment-0 map for the velocity range of [95.0, 106.2] {\kps} (1$\sigma$ = 0.15 K {\kps}). (f) RAMPS NH$_3$(1--1) moment-1 map. 
The rectangular region highlighted in panels ``a,''  ``c,'' and ``e'' is identical to the one shown in Figure~\ref{fg1}. A scale bar of 15 pc is provided in each panel.}

\label{fg2}
\end{figure*}

Figures~\ref{fg2}a and~\ref{fg2}b show the moment-0 map and the intensity-weighted velocity (moment-1) map for the GRS {\grstco} data, respectively. In Figures~\ref{fg2}c and~\ref{fg2}d, we present the moment-0 and moment-1 maps of the CHIMPS {\chmstco} data, respectively. The moment-0 and moment-1 maps for RAMPS NH$_3$(1--1) data are shown in Figures~\ref{fg2}e and~\ref{fg2}f, respectively.
As mentioned in Section~\ref{sec_data_N59}, CHIMPS {\chmstco} and RAMPS NH$_3$(1--1) data have better resolution and trace relatively higher-density gas as compared to GRS {\grstco} data. By combining these data sets, we obtain a comprehensive view of the gas morphology and velocity across diffuse ($\lesssim$ $10^3$ cm$^{-3}$), intermediate ($\sim$ $10^3$ cm$^{-3}$) and denser ($\geq 10^4$ cm$^{-3}$) gas components. Note that Figure~\ref{fg2}a is identical to Figure~\ref{fg1}c; however, it has been included in Figure~\ref{fg2} alongside other panels for ease of comparison. 

The {\chmstco} and NH$_3$(1--1) moment-0 maps reveal the filamentary morphology of the higher-density gas within the extended molecular emission traced in the {\grstco} moment-0 map. Molecular condensations are traced toward the ATLASGAL clumps in both {\chmstco} and NH$_3$(1--1) moment-0 maps. All the moment-1 maps reveal similar velocities at the opposite edges of the filament (see clumps ATL-1, ATL-6, ATL-7). There are significant velocity variations of a few {\kps} along the length of the filament, from its edges toward the center. 
The lowest velocity is observed toward the central region of the filament, where C-HFS or ATL-5 is situated.
Notably, intense molecular emission is detected in each moment-0 maps toward the bubble N59-North, which is associated with ATL-2 (see Figures~\ref{fg2}a, \ref{fg2}c, and \ref{fg2}e). The moment-1 maps indicate that the molecular gas toward the bubble N59-North (including ATL-2) exhibits velocity variations compared to its surrounding areas. This suggests that the massive stars responsible for the bubble N59-North might have influenced the morphology 
and gas kinematics in the filament. This aspect is explored in detail in Section~\ref{pv_ppv_N59}. The study by \citet{chen24} identified {\htwo} regions driven by massive OB-type stars within the filament hosting the N59-North bubble (see Figure 1 in their work). Therefore, this filament can also be examined in the context of the escape and confinement of ionizing radiation from massive stars forming within it \citep[e.g.,][]{whitworth21}.



\subsubsection{Position-velocity and position-position-velocity diagrams using {\grstco} and {\chmstco} data}
\label{pv_ppv_N59}
\begin{figure*}
\center
\includegraphics[width=0.8\textwidth]{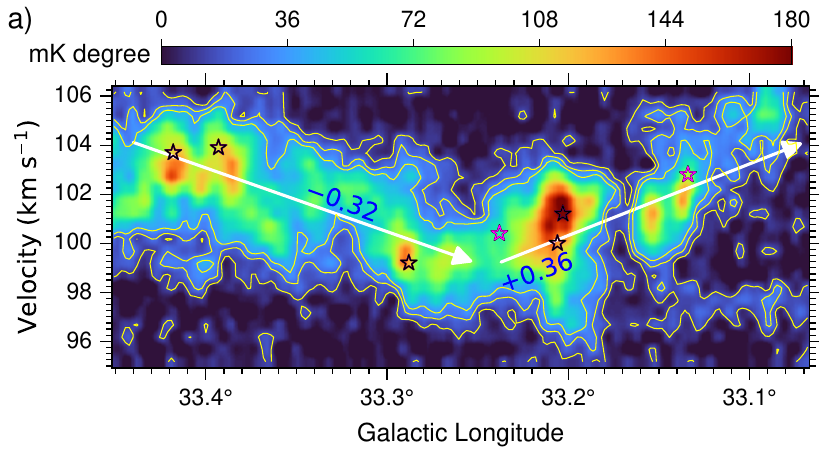}
\includegraphics[width=0.9\textwidth]{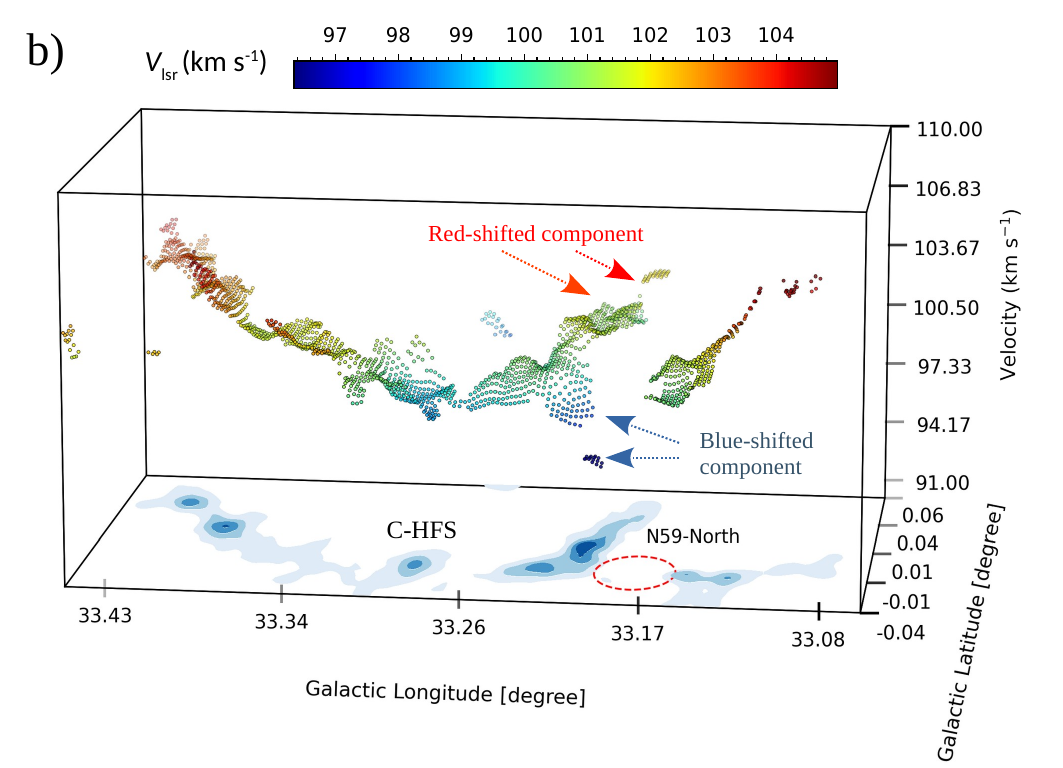}
\caption{(a) The Galactic longitude-velocity (i.e., {\it l-v}) diagram of the {\chmstco} data for the integration range in Galactic latitude = [$-$0$\rlap.{^\circ}$046, 0$\rlap.{^\circ}$059]. The contours are drawn at about [6, 12, 18]$\times \sigma$, where 1$\sigma$ $=$ 2 mK degree. The velocity gradients in the eastern and western parts of the filament are indicated by white arrows, and the corresponding velocity gradient values are provided in the figure in units of {\kps} pc$^{-1}$. (b) The \texttt{SCOUSEPY}-generated position-position-velocity (PPV; here, {\it l-b-v}) diagram. The data points in the diagram represent the position and centroid velocity of the Gaussian components identified in the {\chmstco} emission. The data points are also colored based on their velocity, according to the color scale shown at the top of the image. The {\chmstco} integrated intensity map is shown in the {\it l-b} plane using filled contours at levels [5, 10, 15, 20, 25] K {\kps}. The location of the bubble N59-North is highlighted with a red ellipse in the {\it l-b} plane. The blue- and red-shifted velocity components toward N59-North are indicated with arrows.}
\label{fg4}
\end{figure*}

To reveal the kinematics of the dense gas, we produced the Galactic longitude-velocity (i.e., $l$-$v$ or position-velocity (PV)) diagram using the {\chmstco} data, which is shown in Figure~\ref{fg4}a. The white arrows in Figure~\ref{fg4}a represent the velocity gradients identified in the PV diagram toward the eastern and western parts of the filament. We have estimated these gradients to be about $-0.32$ and $0.36$ {\kps} pc$^{-1}$, respectively. In the case of the $l$-$v$ diagram, the integration range in Galactic latitude is too large (i.e., [$-$0$\rlap.{^\circ}$046, 0$\rlap.{^\circ}$059]) to properly reveal the gas kinematics toward the bubble N59-North. Therefore, we performed the spectral decomposition of the {\chmstco} data using the Python-based tool {\href{https://scousepy.readthedocs.io/en/latest/index.html}{\tt SCOUSEPY}} \citep{henshaw16,henshaw19}. The process began by defining the size of the `Spectral Averaging Areas (SAA)' in pixels, with a selected size of 3 $\times$ 3 pixel$^2$. These SAAs are spatially distributed to cover all emission above a specific threshold ($\sim$1.5 K). From each of these SAAs, an averaged spectrum was extracted and fitted with single or multiple Gaussian components. The best-fit parameters from the SAA-averaged spectra were then utilized to fit the spectra at each pixel in the SAA. To visualize the results, the centroid velocity/velocities of the fitted Gaussian(s) is/are plotted for each pixel in position-position-velocity (PPV; here, {\it l-b-v}) space. The resulting PPV diagram from our analysis is shown in Figure~\ref{fg4}b, where the moment-0 map of the filament is displayed using filled contours in the $l$-$b$ plane. The PPV map is consistent with the PV diagram, revealing opposite velocity gradients toward the eastern and western parts of the filament. Notably, the PPV map clearly highlights the red- and blue-shifted gas components toward the bubble N59-North (see the arrows).
\subsection{Star fomation activities in the filamentary cloud}
\label{star_N59}
\begin{figure*}
\center
\includegraphics[width=13.8 cm]{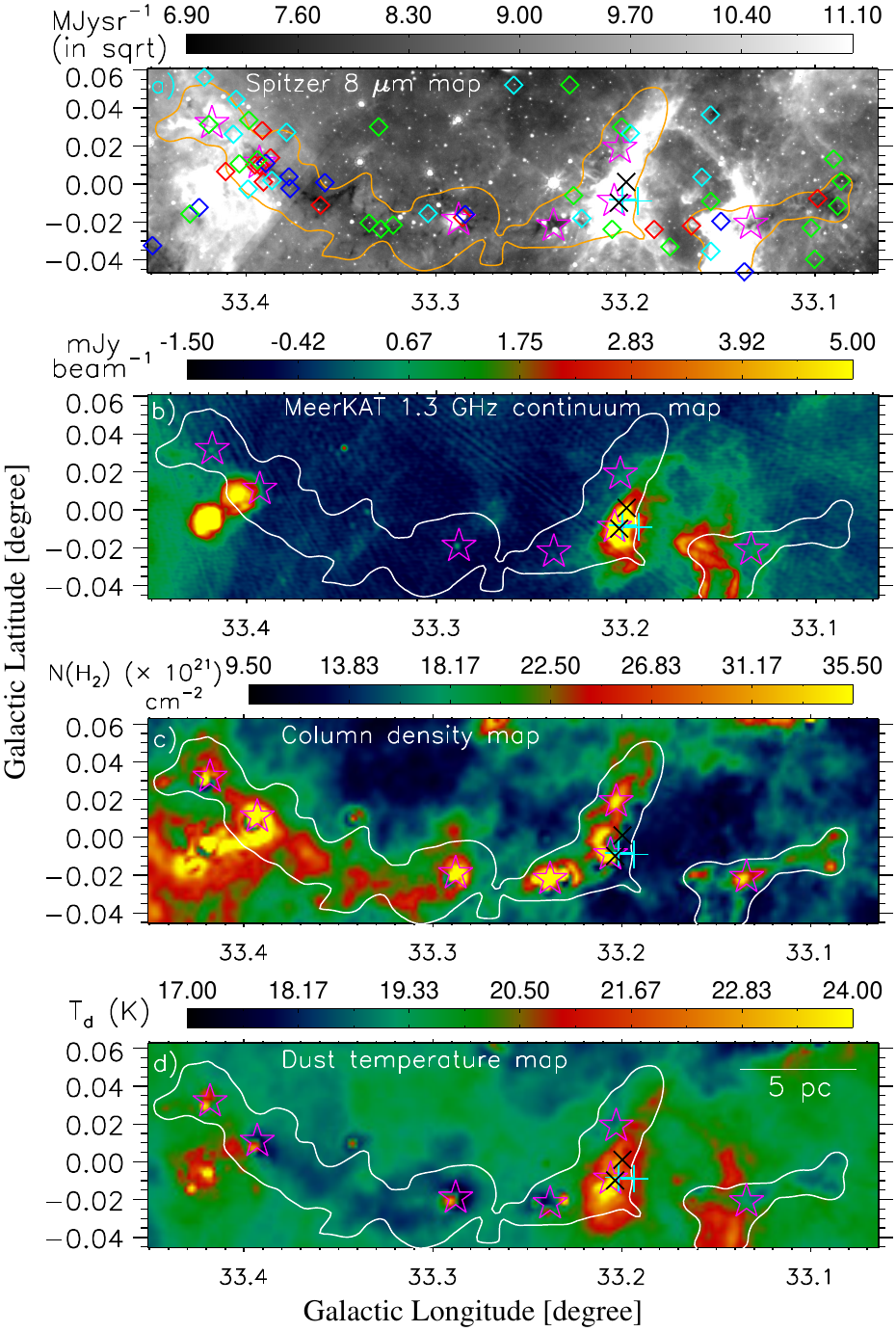}
\caption{(a) {\it Spitzer} 8 $\mu$m image, (b) MeerKAT 1.3 GHz radio continuum emission map, (c) H$_2$ column density ($N$(H$_2$)) map, and
(d) Dust temperature ($T_{\rm{d}}$) map, corresponding to the region outlined by red dotted-dashed lines in Figures~\ref{fg1} and \ref{fg2}. In each panel, the {\chmstco} emission is shown using a contour at about 3.7 K {\kps}. The magenta star, black cross, and cyan plus symbols indicate the position of the ATLASGAL clumps, 6.7 GHz MMEs, and UC {\htwo} regions, respectively. The YSO candidates are shown using diamond symbols in panel ``a.'' The Class I, Flat Spectrum, and Class II sources obtained from the SPICY catalog \citep{kuhn21} are colored red, cyan, and green, respectively. The blue diamonds are additional Class I YSO candidates satisfying the color conditions: [4.5] $-$ [5.8] $\geq$ 0.7 and [3.6] $-$ [4.5] $\geq$ 0.7 \citep{getman_2007}.  A scale bar of 5 pc is marked in panel ``d.''}
\label{fg3}
\end{figure*}

Figures~\ref{fg3}a,~\ref{fg3}b,~\ref{fg3}c, and~\ref{fg3}d present overlays of the locations of the ATLASGAL clumps and the {\chmstco} emission contour at about 3.7 K {\kps} (i.e., {\simi}10$\sigma$) on four different maps: the {\it Spitzer} 8.0 $\mu$m image, the MeerKAT 1.3 GHz continuum emission map, the {\it Herschel} H$_2$ column density map, and the {\it Herschel} dust temperature map, respectively. These maps help us to identify the presence of dense molecular gas, dust clumps, ionized emission, cold dust, warm dust, and column density toward the filamentary cloud. The filamentary cloud is linked to material exhibiting column densities exceeding 
2.25 $\times$ 10$^{22}$ cm$^{-2}$, and is assoicated with warm dust ($T_\mathrm{d}$ = 20.5--24~K) as well as relatively cold dust emission ($T_\mathrm{d}$ = 17--18.5~K). All the ATLASGAL clumps are seen toward the areas where column densities exceed 3.2 $\times$ 10$^{22}$ cm$^{-2}$. 
The radio continuum emission map reveals the association of extended ionized regions with the filamentary cloud, where the presence of warm dust emission is evident. In the filamentary cloud, dust emission at $T_\mathrm{d}$ = 17--18.5~K  is traced in regions within the longitude range of [$33\rlap.{^\circ}25$--$33\rlap.{^\circ}40$]. 

The detection of YSOs within a molecular cloud provides as direct evidence of ongoing star formation. The IR excess observed in YSOs arises from their envelopes and dusty circumstellar disks \citep{Sharma_2017MNRAS,kuhn21}. To analyze the distribution of  YSOs within the filamentary structure hosting N59-North (outlined by the magenta dot-dashed lines in Figure~\ref{fg1}a), we used the SPICY catalog by \citet{kuhn21}. In the {\it Spitzer} 8.0 $\mu$m image, the positions of Class~I, Flat Spectrum, and Class~II YSOs from the catalog are highlighed by red, cyan, and green diamonds, respectively (see Figure~\ref{fg3}a). 
Using the color conditions [4.5] $-$ [5.8] $\geq$ 0.7 and [3.6] $-$ [4.5] $\geq$ 0.7 \citep{getman_2007}, we have also identified additional Class~I YSO candidates, represented by blue diamonds in Figure~\ref{fg3}a. These YSOs are distinct from those reported in \citet{kuhn21}. Overall, numerous YSO candidates are detected within the {\chmstco} moment-0 contour, indicating the association of higher-density gas with ongoing star-forming activity. 
%
%
\section{Discussion}
\label{sec_disc_N59}
Through a comprehensive multi-wavelength observational investigation, we have gained new insights into the MSF activity toward the bubble N59-North. Our findings include the identification of an extended IRDC/filamentary structure that exhibits significant velocity variation along its length. Additionally, we have revealed several dust clumps as potential candidates for MSF and identified a C-HFS. These results are discussed in this section to infer the ongoing physical processes related to the origin, evolution, and MSF activity of the IRDC hosting the bubble N59-North.

\subsection{The massive star-forming activity in the filamentary cloud}
Massive stars form inside hot molecular cores \citep[][]{Mayra_1999,Tak_2004,Paron_2024}. Later, the intense ultraviolet radiation from massive stars, beyond the Lyman limit, ionizes the surrounding gas \citep[][]{Panagia_1973}. The size of {\htwo} regions increases over time; thus, the compact {\htwo} regions ($<0.5$ pc) are associated with the early stages of MSF. In this study, we have detected a pc-scale HFS candidate (i.e., C-HFS) at the center of the filament, which is associated with ATL-5 and an UC {\htwo} region. The presence of the UC {\htwo} region and the absence of any extended {\htwo} regions suggest that C-HFS represents an early stage of HFS. Assuming that the {\htwo} region expanding in a uniform medium, we calculated the dynamical age for the UC {\htwo} region to be about 0.01--0.04 Myr for the initial densities $n = 10^4$ to $10^5$~cm$^{-3}$, respectively, using the formula from \citet{Dyson_1980}. In this calculation, the sound speed in the ionized region  \citep[$\sim$10 km s$^{-1}$;][]{Bisbas_2009}, as well as its effective radius ({\simi}0.05 pc) and Strömgren radius, are provided as inputs. For densities ranging from $n = 10^4$ to $10^5$~cm$^{-3}$, the Strömgren radius of the {\htwo} region varies from 0.006 to 0.001 pc. 
It is possible that the target IRDC hosts several UC {\htwo} regions and 6.7 GHz MMEs toward the edge of the N59-North bubble. Recently, \citet{chen24} calculated the dynamical timescale for the extended {\htwo} region associated with N59-North to be about 2 Myr and proposed that the expansion of the {\htwo} is responsible for the star-forming activity near N59-North. This study supports their proposed idea by revealing the expanding gas motion near the N59-North bubble by detecting blue- and red-shifted velocity components. Therefore, taking all the information together, we find that our target IRDC has been active in MSF for the last few Myr.

The presence of several pc-scale filaments associated with C-HFS at the central part of the large-scale filament indicates the importance of multiple scales filamentary mass accretion related to MSF as proposed by \citet{Zhou_2022MNRAS}. The gas flow in the filaments on small scale is primarily driven by gravity, whereas the large-scale gas flow is possibly caused by both turbulence (as inertial flow) and gravitational contraction \citep[e.g.,][]{Zhou_2022MNRAS,Bhadari_2025_A&AL}. Thus, the hubs accumulate a large amount of material and become suitable for MSF. The observed large-scale velocity gradients toward the filament hosting N59-North possibly suggest a converging gas flow, which is providing material to C-HFS \citep[e.g.,][]{kirk_2013}. More details about the large-scale gas motion and its possible origin are given in Section~\ref{origin_N59}. The GRS and CHIMPS line data have insufficient resolution to trace C-HFS; therefore, small-scale gas motion remains unexplored in this work. It is important to note that although all ATLASGAL clumps satisfy the condition for MSF, we have not observed pc-scale HFS systems associated with them, except for ATL-5. This suggests that, in the case of these clumps, the HFS configuration may arise at later stages or may be too weak to be detected with the current sensitivity of the data. As demonstrated by the recent observational work by \citet{Bhadari_2025_A&AL}, a combination of ALMA and JWST observations can be extremely useful for hunting such early stages of massive star-forming regions and understanding the driving mechanisms of gas flow in small-scale.

\subsection{The possible origin and evolution of the filamentary cloud} \label{origin_N59}
\begin{figure*}
\center
\includegraphics[width=\textwidth]{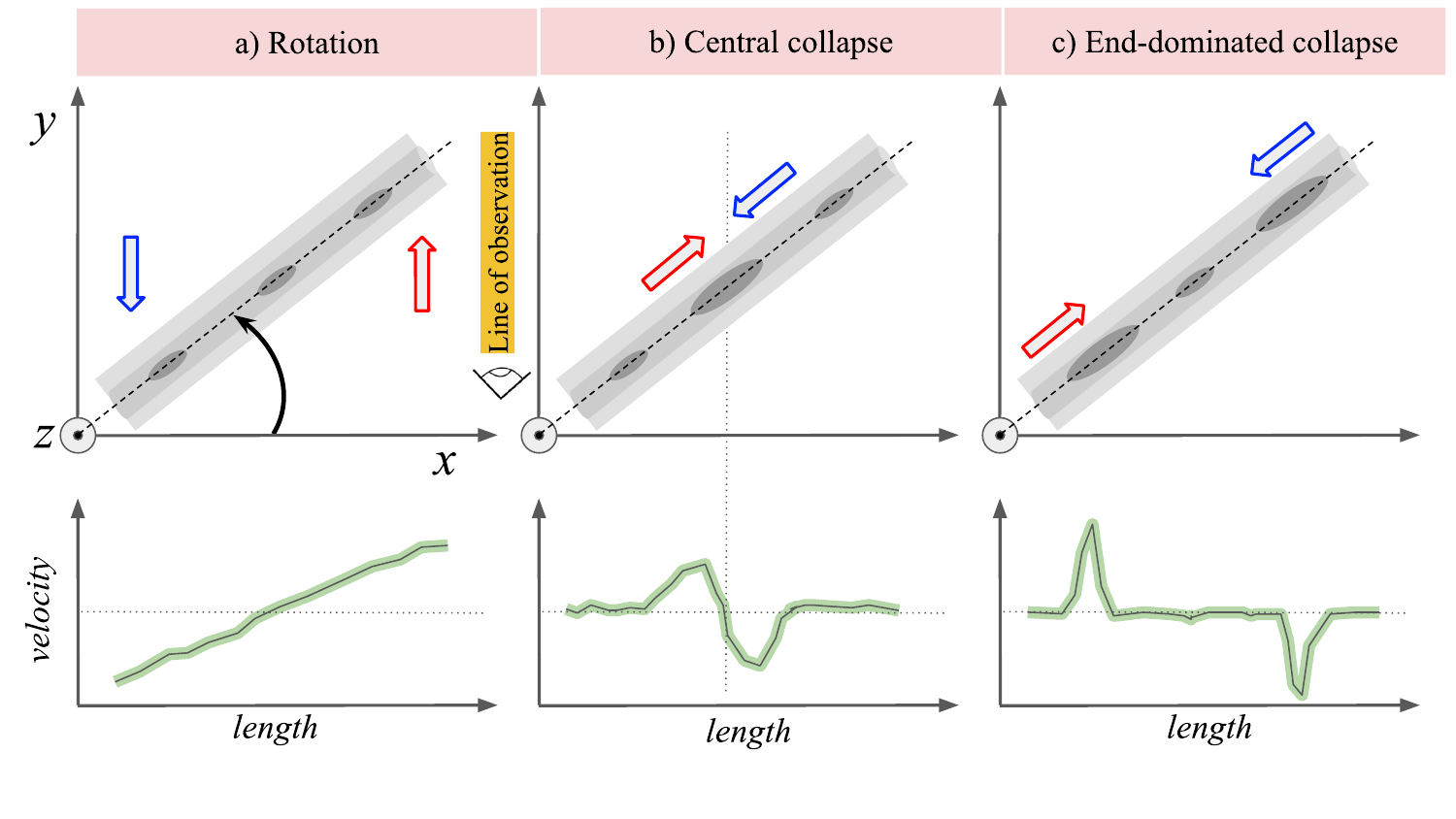}
\caption{A schematic view of the possible position-velocity (PV) diagrams for cylindrical filaments in different gas motions. The $x$-$z$ plane represents the plane of the sky, and the $y$-direction corresponds to the line of observation, which forms a significant angle ($\sim 45\deg$) relative to the filament. (a) A rotating filament with a constant angular velocity, indicated by a curved arrow. (b) A filament undergoing central collapse \citep[motivated from][]{Liu_2019}. (c) A filament undergoing end-dominated collapse \citep[motivated from][]{Clarke_2015}. The blue and red arrows in each panels represent the blue- and red-shifted velocity components. 
} 
\label{fg5}
\end{figure*}

The PV/PPV diagrams show significant velocity variation ($\sim$ few {\kps}) along the filament, which can be very important for inferring the possible mechanisms driving the large-scale gas motion. In Figures~\ref{fg5}a, \ref{fg5}b, and \ref{fg5}c, we provide schematic diagrams explaining how the PV diagrams would appear under different processes of gas flow, such as rotation, central collapse, and end-dominated collapse (EDC), respectively.
The PV diagrams are shown for cylindrical filaments and line of observation making significant angle ($\sim45\deg$) relative to the filament. For a rotating filament with constant angular velocity, the observed velocity will be proportional to the distance along the spine of the filament. 
A centrally collapsing filament will develop a massive clump toward its central region, exhibiting both blue- and red-shifted velocity components directed toward the filament's center. This kinematic signature can be inferred from the study by \citet{Liu_2019}. According to \citet{Clarke_2015}, due to varying gravitational acceleration along the length of the isolated filaments, they undergo EDC. In the EDC process, massive clumps are exclusively found at the opposite edges of the filament, having blue- and red-shifted velocity components. It is important to note that if the filament is parallel to the line of observation, its filamentary nature will not be observed. Whereas, if the filament is perpendicular to the line of observation, blue- or red-shifted components will only be visible for rotation. The PV/PPV diagrams of the filament hosting N59-North do not correspond to any of the PV diagrams shown in Figure~\ref{fg5}. Therefore, rotation, central collapse, and EDC are unlikely to be responsible for the large-scale gas flow in this filament.

\begin{figure*}
\center
\includegraphics[width=0.85\textwidth, trim={0cm 2cm 0cm 1cm}, clip]{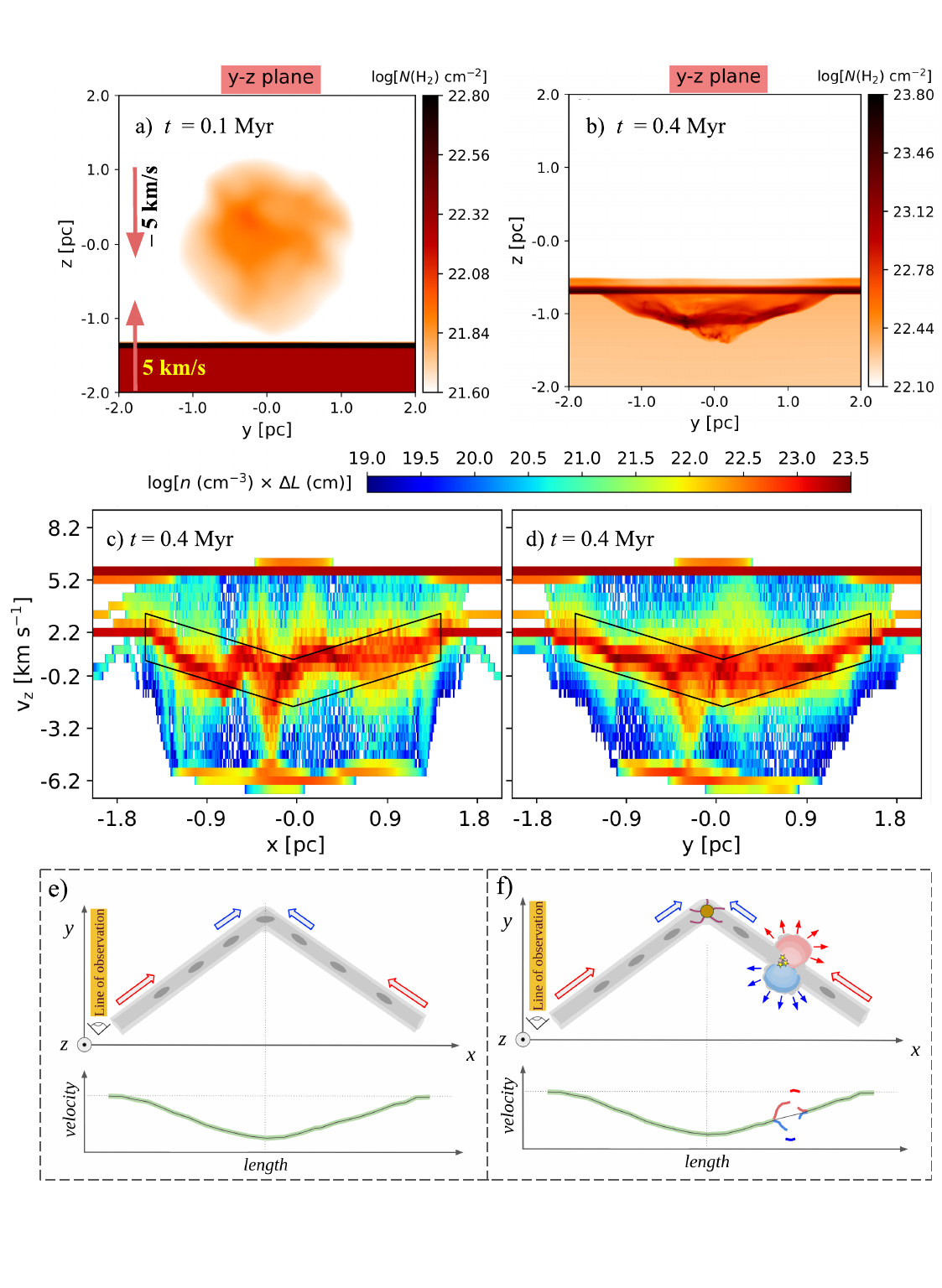}
\caption{(a) The $N$(H$_2$) map in the $y$-$z$ plane at 0.1 Myr, showing the collision of a turbulent molecular cloud (radius = 1.5 pc) and a sea of dense gas moving with a relative velocity of 10 {\kps} along the $z$-direction. (b) The $N$(H$_2$) map in the $y$-$z$ plane at 0.4 Myr. Panels (c) and (d) present the PV diagrams along the $x$- and $y$-axes, respectively, for a width of 0.4 pc. The area enclosed by the black lines indicates the velocity of the compressed gas. These figures are taken from \citet{Maity_2024arXiv}. The schematic diagrams in panels (e) and (f) depict the initial and current configurations of our target filament in position and velocity space, respectively. The current configuration shows the presence of a HFS at the center of the filament, bubble N59-North, and its role in creating blue- and red-shifted velocity components. Stars indicate the presence of radio continuum sources toward N59-North. Similar to Figure~\ref{fg5}, the $x$-$z$ plane represents the plane of the sky, while the $y$-direction corresponds to the line of observation. The blue and red arrows represent the blue- and red-shifted velocity components in the filament. Note that the schematic figures are not to scale.} 
\label{fg6}
\end{figure*}

Interestingly, the PV diagram for the target filament matches perfectly with that of the compressed layer observed in the cloud-cloud collision (CCC) scenario presented in \citet{Maity_2024arXiv} using magneto-hydrodynamic (MHD) simulation data from \citet{inoue18}. The V-shaped PV structure was first analyzed using the same data by \citet{Arzoumanian_2018PASJ} (see Figure 16 in their work). The CCC model of \citet{inoue18} involves the collision of a turbulent molecular cloud (radius of 1.5 pc) and a dense sea of gas, moving at a relative velocity of 10 {\kps}, as shown in Figure~\ref{fg6}a. This CCC model represents a case where a larger cloud collides with a smaller one or a cloud is subjected to compression by a plane-parallel shock wave. More details about the numerical simulation can be found in \citet{inoue18} and \citet{Maity_2024arXiv}. The collision resulted in a cone-shaped compressed layer as presented in Figure~\ref{fg6}b at 0.4 Myr. Since molecular clouds are not ideal spheres in reality, the resulting cones can be imperfect or distorted, influenced by the molecular cloud's initial structure. Figures~\ref{fg6}c and \ref{fg6}d show the PV diagrams at 0.4 Myr along the $x$ and $y$ axes for the line of observation toward $z$-direction, respectively. The area covered by the black lines highlights the compressed layer in the PV diagrams. The opposite signs of the velocity gradients in the left and right part of the PV diagrams indicate converging gas flow toward the vertex of the cone. The similarity of the observed PV diagram for the filament hosting N59-North and the PV diagrams for a converging gas flow in a cone suggests that the filament must have a cone-like shape. This indicates that the gas is converging toward the center of the filament, where C-HFS is situated. Figure~\ref{fg6}e presents a schematic diagram that illustrates the cone-like shape of the filament and the variation in velocity along its length before the emergence of the N59-North bubble. The converging gas flows toward the center of the filament give rise to C-HFS. The emergence and expansion of bubble N59-North, along with its effect on the PV diagram, is illustrated in Figure~\ref{fg6}f.

\begin{figure*}
\center
\includegraphics[width=0.50\textwidth]{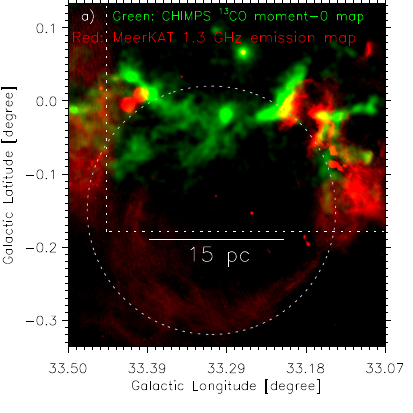}
\includegraphics[width=0.50\textwidth]{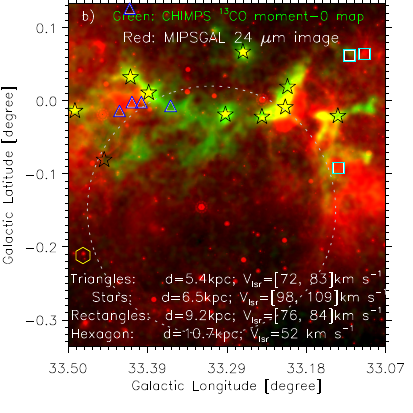}
\caption{(a) A large-scale view of our target site is shown using a two-color composite image. Colors red and green present the MeerKAT 1.3 GHz radio continuum emission and {\chmstco} moment-0 map, respectively. The radio continuum emission is displayed on a linear scale from 5$\sigma$ to 100$\sigma$, where 1$\sigma$ $\sim$ 20 $\mu$Jy beam$^{-1}$. The moment-0 map is shown on a linear scale from 3$\sigma$ to 30$\sigma$, where 1$\sigma$ $\sim$ 0.36 K {\kps}. The white dotted rectangle indicates the area shown in Figure~\ref{fg1}a. A scale bar of 15 pc is marked in this panel. (b) This panel presents a similar two-color composite image as panel ``a,'' with red representing the MIPSGAL 24 µm image. The MIPSGAL image is displayed on a logarithmic scale, ranging from 25 to 200 MJy sr\(^{-1}\). The distribution of the ATLASGAL clumps for different velocity ranges and distances is indicated using various symbols, as described in the figure. The extent of a candidate {\htwo} region from \citet{Anderson_2014ApJS} is marked with a dotted circle in each panel.} 
\label{fg7}
\end{figure*}

As mentioned earlier, the model proposed by \citet{inoue18} is also applicable in the case of a molecular cloud interacting with a plane-parallel shock front arising from the expansion of an {\htwo} region. We examined MeerKAT 1.3 GHz radio continuum data and MIPSGAL 24 $\mu$m data for an extended region to investigate this possibility. An extended {\htwo} region is evident in both the MeerKAT and MIPSGAL data, as shown in Figures~\ref{fg7}a and \ref{fg7}b, respectively. This {\htwo} region is classified as a candidate {\htwo} region in \citet{Anderson_2014ApJS} and is highlighted with a dotted circle in Figure~\ref{fg7}. However, our target filament extends beyond the boundaries of the {\htwo} region, and molecular gas is not traced at the edge of this {\htwo} region in the velocity range of the filament. Therefore, the {\htwo} region is possibly not associated with the filament.

\subsection{The review of CCC scenario toward N59-North}
As mentioned in Section~\ref{sec:intro}, a CCC event was reported toward this target site by \citet{chen24}. They proposed a collision of two cloud components with velocities of [65, 79] and [95, 108] {\kps}, having a relative velocity of about 24 {\kps}. We found different distances for the ATLASGAL clumps with the velocity ranges of [72, 83] and [98, 109] {\kps}. The distribution of the ATLASGAL clumps according to their velocities and distances is shown using different symbols in Figure~\ref{fg7}b. It is important to note that, the histogram of the Galactic CCC events based on their collision velocities shows that the number of CCC events decreases above 5 {\kps} and collision with a relative velocity of 24 {\kps} is extremely rare \citep{fukui21}. In addition, the origin of the entire filament hosting N59-North is not established based on the collision of the above-mentioned velocity components. Therefore, considering all these points, it is unlikely that the collision of velocity components [65, 79] and [95, 108] {\kps} is responsible for the formation of the filament hosting N59-North and its massive star-forming activity.

We can enhance the understanding of the possible CCC event at this target site based on the study of \cite{Maity_2024arXiv}. No clear signature of two velocity components is observed in the range [95, 108] {\kps} (see Figure~\ref{fg4}) for the entire filament. As detailed in \cite{Maity_2024arXiv}, the absence of two velocity components in CCC sites is very much possible as the signature of the two velocity components, and their connection \citep[known as bridge feature;][]{torri_2011,fukui_2014,fukui_2015,dewangan_2017,Sano_2018PASJ,fujita21} remains for a very short period. From the beginning of the collision, the timescale is limited by the size of the cloud components and their relative velocity. Beyond that timescale, it becomes impossible to separate the individual velocity components. Therefore, the possibility of CCC toward this target site remains; however, the exact determination of the colliding cloud components within the velocity range of [95, 108] {\kps} is not feasible with our current understanding of CCC. Such analysis would require new methods.

Several studies based on MHD simulations \citep[e.g.,][]{inoue13,inoue18,Fukui_2021PASJ}, have demonstrated that the formation of high-density filaments is efficient perpendicular to the magnetic field. While in the case of CCC, the filaments can form parallel to the magnetic field, however, they disperse over time \citet{Maity_2024arXiv}. Therefore, high-resolution dust polarization observations will be helpful in examining the magnetic field morphology for the filament hosting N59-North. Interestingly, the formation and distribution of the massive dense cores suitable for MSF depend upon the inhomogeneous density structures of the colliding molecular clouds before the collision. Thus, the massive star-forming activity is not strictly restricted to the vertex part of the cone/central part of the filament only. The possibility of MSF, apart from the central part of the filament, accounts for the formation of stars which are responsible for the bubble N59-North.

\section{Summary and Conclusions}
\label{sec_conc_N59}
To understand the physical processes related to MSF, we conducted a multi-wavelength observational investigation for an extended area hosting N59-North. The major outcomes of this study are summarized below. 

\begin{enumerate} 
	\item An elongated IRDC (length {\simi}28 pc) is investigated in the {\it Spitzer} 8 $\mu$m image, which is not reported in the literature. This IRDC hosts bubble N59-North, multiple protostars, and seven ATLASGAL dust clumps at the same distance.
		
	\item The GRS {\grstco}, CHIMPS {\chmstco} and RAMPS NH$_3$(1--1) line data confirm the existence of this elongated filamentary structure, which is traced in a velocity range of about [95, 106] km s$^{-1}$.

	\item All ATLASGAL clumps meet the emperical mKP-10 criteria \citep[i.e., $M(R) > 580$ {\Msolar} $({R}/{\mathrm{pc}})^{1.33}$;][]{urquhart18} for MSF. 
	
	\item Using the {\it Spitzer} 8 $\mu$m image, a new C-HFS is investigated toward the ATLASGAL clump located at the central part of the filament. In the direction of C-HFS, we have detected an UC {\htwo} region driven by a B2-type star based on MeerKAT 1.3 GHz continuum emission. The lack of extended ionized emission toward C-HFS suggests that it is in the early evolutionary stage with no significant feedback from the young massive star. 
	
		
	\item The study of the observed velocity features in the CHIMPS {\chmstco} PV/PPV diagrams toward the filament and the existing theoretical models shows that physical processes such as rotation, central collapse, or EDC are not responsible for the observed gas motion in the filament. Additionally, the comparison of PV/PPV diagrams with the results of MHD simulations \citep{inoue18,Maity_2024arXiv} suggests that the filament is possibly conical in shape and exhibits converging gas motion toward its center.

	\item The blue- and red-shifted gas velocities observed at the edges of the bubble N59-North in the PPV diagram show an expanding gas motion.

	\item The outcomes of this study favor CCC activity in the filament at [95, 106] km s$^{-1}$, but they contradict previous claims of a collision between the [65, 79] and [95, 108] km s$^{-1}$ components, as they are at different distances. Although the exact colliding components remain unidentified, comparisons of the outcomes of this study with MHD simulations suggest that CCC contributed to the filament’s formation and gas motion.


\end{enumerate}
Taken together all the results, the filament hosting N59-North displays a converging gas motion toward its center, where a pc-scale C-HFS is detected at an early evolutionary stage of HFSs. The converging flow toward C-HFS supports the idea of multi-scale filamentary mass accretion for MSF, likely triggered by CCC.
Notably, the astrometric distance estimation using Gaia parallax measurements and the measured distance of the ATLASGAL dust clumps suggest that our target filamentary cloud could be located at either 4.66 kpc or 6.5 kpc. Therefore, further investigation is necessary to determine the precise distance of this filamentary cloud.


%
\section{Data availability}
\label{sec_data_avail}
The {\it Herschel} and {\it Spitzer} data are available in the publicly accessible \href{https://irsa.ipac.caltech.edu/frontpage/}{NASA/IPAC Infrared Science Archive}. The SPICY catalogue can be downloaded from the \href{https://irsa.ipac.caltech.edu/cgi-bin/Gator/nph-scan?projshort=Spitzer&mission=irsa}{{\it Spitzer} Catalogue}. The MeerKAT 1.3 GHz radio continuum data is available in \href{https://archive-gw-1.kat.ac.za/public/repository/10.48479/3wfd-e270/index.html}{SMGPS Data Archive}, while the 20 cm radio continuum data can be accessed via the \href{https://third.ucllnl.org/gps/}{MAGPIS Data Archive}. The {\it Herschel} column density and dust temperature maps are available in the \href{http://www.astro.cardiff.ac.uk/research/ViaLactea/}{ViaLactea Data Server}. The GRS $^{13}$CO($J = $ 1--0) datacubes are available in \href{https://www.bu.edu/galacticring/new_data.html}{Boston University Data Archive}, and the CHIMPS $^{13}$CO($J = $ 3--2) data can be found in the \href{https://www.canfar.net/storage/vault/list/AstroDataCitationDOI/CISTI.CANFAR/16.0001/data/CUBES/13CO}{CANFAR Archive}. The RAMPS NH$_3$(1--1) data are available in the \href{https://greenbankobservatory.org/category/portal/gbt/gbt-legacy-archive/}{GBT Legacy Archive}. ATLASGAL data is accessible via the \href{https://atlasgal.mpifr-bonn.mpg.de/cgi-bin/ATLASGAL_DATABASE.cgi}{ATLASGAL Database}, and the physical parameters of the ATLASGAL clumps are catalogued in the \href{https://vizier.cds.unistra.fr/viz-bin/VizieR-3?-source=J/MNRAS/473/1059/}{VizieR Data Archive}.

\section*{Acknowledgments}
We thank the anonymous referee for insightful comments and suggestions, which have enhanced the scientific quality of this paper. The research work at Physical Research Laboratory is funded by the Department of Space, Government of India. This work is based [in part] on observations made with the {\it Spitzer} Space Telescope, which is operated by the Jet Propulsion Laboratory, California Institute of Technology, under a contract with NASA. The MeerKAT telescope is operated by the South African Radio Astronomy Observatory, which is a facility of the National Research Foundation, an agency of the Department of Science and Innovation. This research has made use of the VizieR catalogue access tool, CDS, Strasbourg, France (DOI : 10.26093/cds/vizier). The original description of the VizieR service was published in 2000, A\&AS 143, 23. This publication makes use of molecular line data from the Radio Ammonia Mid-Plane Survey (RAMPS). RAMPS is supported by the National Science Foundation under grant AST-1616635. This publication makes use of the GBT Legacy Archive. This research made use of \href{http://www.astropy.org}{{\it Astropy}}, a community-developed core Python package for Astronomy \citep{astropy13,astropy18}. For figures, we have used {\it matplotlib} \citep{Hunter_2007} and IDL software.


\newpage

%



\bibliographystyle{aasjournal}
\bibliography{N59_AKM_R1}{}

\appendix
\restartappendixnumbering
\section{Physical parameters of the ATLASGAL clumps} \label{sec_appendix_tau13}
The physical parameters of the ATLASGAL clumps are provided in Table~\ref{tab_atlasgal_N59}.


\begin{table*}
\centering
\caption{The physical properties of the ATLASGAL clumps. The velocity, mass, and effective radius of these clumps are obtained from \citet{urquhart18}. The mass and effective radius values presented in the table have been rescaled to the adopted distance of {\it d} = 4.66 kpc. Information on outflow activity associated with these clumps is sourced from \citet{Yang2018ApJS}.}
\begin{tabular}{ccccccc}
\hline \hline
 ID &      {\it l} &      {\it b} &     {\it v} &        $R_{\rm{eff}}$ &        log($M$ [{\Msolar}]) & Outflow \\
    &        (deg) &        (deg) &    ({\kps}) &                  (pc) &                             &  \\
 \hline
  1 & 33.134 & -0.021 & 102.8 & 0.22 & 2.46 &      No \\
  2 & 33.203 &  0.019 & 101.2 & 0.50 & 2.98 &     Yes \\
  3 & 33.206 & -0.009 & 100.0 & 0.90 & 2.95 &     Yes \\
  4 & 33.238 & -0.022 & 100.4 & 0.88 & 3.09 &      -- \\
  5 & 33.288 & -0.019 &  99.2 & 0.34 & 2.40 &     Yes \\
  6 & 33.393 &  0.011 & 103.9 & 1.02 & 3.38 &     Yes \\
  7 & 33.418 &  0.032 & 103.7 & 0.20 & 2.32 &     Yes \\
\hline \hline
\end{tabular}
\label{tab_atlasgal_N59}
\end{table*}

\end{document}